\documentclass[10pt]{an}
\usepackage{times}
\usepackage{graphicx}

\begin{document}

\Pagespan{0}{1}
\Yearpublication{2015}
\Yearsubmission{2014}
\Month{1}
\Volume{336}
\Issue{1}
\DOI{DOI}

\title{Comparing modal noise and FRD of circular and non-circular cross-section fibres}

\author{D. P. Sablowski\thanks{Corresponding author.
\email{dsablowski@aip.de}}, D. Pl\"uschke, M. Weber, K. G. Strassmeier \and A. J\"arvinen}

\titlerunning{Modal noise and FRD of non-circular and circular fibres}
\authorrunning{D.~P.~Sablowski et al.}

\institute{Leibniz-Institute for Astrophysics Potsdam (AIP), An der Sternwarte 16, 14482 Potsdam,
Germany}

\received{R-date} \accepted{A-date}

\publonline{2015}


\abstract{Modal noise is a common source of noise introduced to the measurements by optical fibres and is particularly important for fibre-fed spectroscopic instruments, especially for high-resolution measurements. This noise source can limit the signal-to-noise ratio and jeopardize photon-noise limited data. 
The subject of the present work is to compare measurements of modal noise and focal-ratio degradation (FRD) for several commonly-used fibres. We study the influence of a simple mechanical scrambling method (excenter) on both FRD and modal noise. Measurements are performed with circular and octagonal fibres from Polymicro Technology (FBP-Series) with diameters of 100, 200 and 300\,$\mu$m and for square and rectangular fibres from CeramOptec, among others. 
FRD measurements for the same sample of fibres are performed as a function of wavelength. Furthermore, we replaced the circular fibre of the STELLA-\'echelle-spectrograph (SES) in Tenerife with an octagonal and found a SNR increase by a factor of 1.6 at 678 nm.
It is shown in the laboratory that an excenter with a large amplitude and low frequency will not influence the FRD but will reduce modal noise rather effectively by up to 180\%. }

\maketitle

\sloppy


\section{Introduction}

High spectral-resolution \'echelle spectrographs have moved to center stage for planet detection and now also for follow-up observations of space-based transit detections (e.g., Pepe et al.~\cite{espresso}). Because such spectrographs must be stabilized to extreme precision to reach ms$^{-1}$  stability, their physical location is generally not anymore at the telescope focus but in an isolated room aside or below the telescope. This requires either an expensive coud\'e train or a cheaper and more-efficient fibre feed. Therefore, the fibre products from the telecommunication industry are very welcomed in astronomy but must be characterized individually due to  the limitations imposed by the extremely low light levels typical for astronomy. In this paper, we present characterizations of a series of different types of fibres with particular focus on modal noise and focal-ratio degradation.
Our direct aim is to compare the fibres of the new high-resolution spectrograph PEPSI for the Large Binocular Telescope (Strassmeier et al.~\cite{pepsi}) with other solutions.       

A study of the literature on modal noise shows that the generality of these measurements is quite limited due to the fact, that these studies are limited to a given instrument. E.g. Grupp~(\cite{grupp}) and Baudrand \& Walker~(\cite{bau:wal}) showed a method how to reduce that noise source in their FOCES and SHARMOR measurements, respectively. Designing a new instrument, however, makes it necessary to know how modal noise depends on the fibre and the fibre coupling system without limitations to a given instrument the fibre is connected to. A short review of common methods to reduce that source of noise is given by Reynolds \& Kost~(\cite{rey:kos}).

Due to interference inside a waveguide, a homogeneous input light beam is transformed to an (angle-) discrete output beam. Waves, which fulfill the condition for total internal reflection (TIR), 
\begin{equation}\label{Eq1}
  \sin{(\theta_{c})}=\frac{n_{2}}{n_{1}}
\end{equation}
at the interface between a core with index of refraction $n_1$ and a cladding with $n_2$ and the condition for self-consistency, are called modes of the waveguide. A self-consistent wave is such a wave, which is reproduced in amplitude and phase after the second TIR.

\begin{figure}[h!]
\begin{center}
\includegraphics[width=85mm]{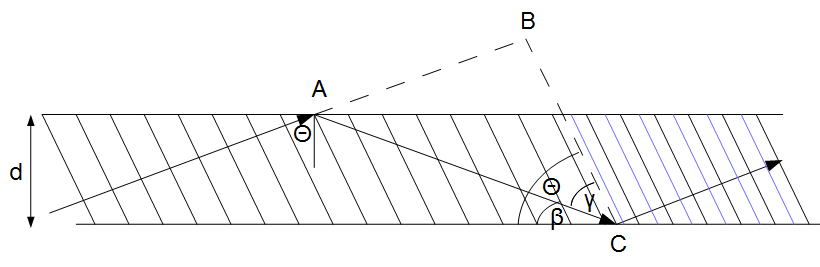}
\end{center}
\caption{Interference in the core of a waveguide. A wave which is reproduced in amplitude and phase after the second TIR  at the interface is called a mode of the waveguide. The energy is transported by these modes and the energy of waves which are not reproduced will be displaced  to the modes of the waveguide by interference.}
  \label{F1}
\end{figure}

From Fig.~\ref{F1} one can extract a condition for constructive interference between the incoming and the second-time reflected wave at point C and finds
\begin{equation}
\label{Eq2}
  m\lambda = 2 d n_{1}\cos{(\Theta_{m})}-\frac{\phi\lambda}{\pi},
\end{equation}
where the second term cares for the phase-shift at reflection. The discrete character of the modes is directly visible from this equation. Furthermore, the squared normalized frequency, $V^{2}$, or waveguide parameter, 
\begin{equation}
\label{Eq3}
  V^{2} =k^{2}\rho^{2}(n_{1}^{2}-n_{2}^{2}) ,
\end{equation}
where $k=2\pi / \lambda$ is the wavenumber and $\rho$ the radius of the fibre-core, is a measure of the numbers of guided modes $N_{\rm M}$ inside a waveguide, e.g., Bures (\cite{bures}) ($V^2 \propto N_{M}$). By using the definition for the numerical aperture, $(NA)^{2}=n_{1}^{2}-n_{2}^{2}$, of a waveguide, Eq.~(\ref{Eq3}) can be rewritten as
\begin{equation}
\label{Eq4}
  V^{2} = \frac{4\pi^{2}}{\lambda^{2}} (NA)^{2}\rho^{2} \ .
\end{equation}

\begin{figure}[h!]
 \begin{center}
 \includegraphics[width=85mm]{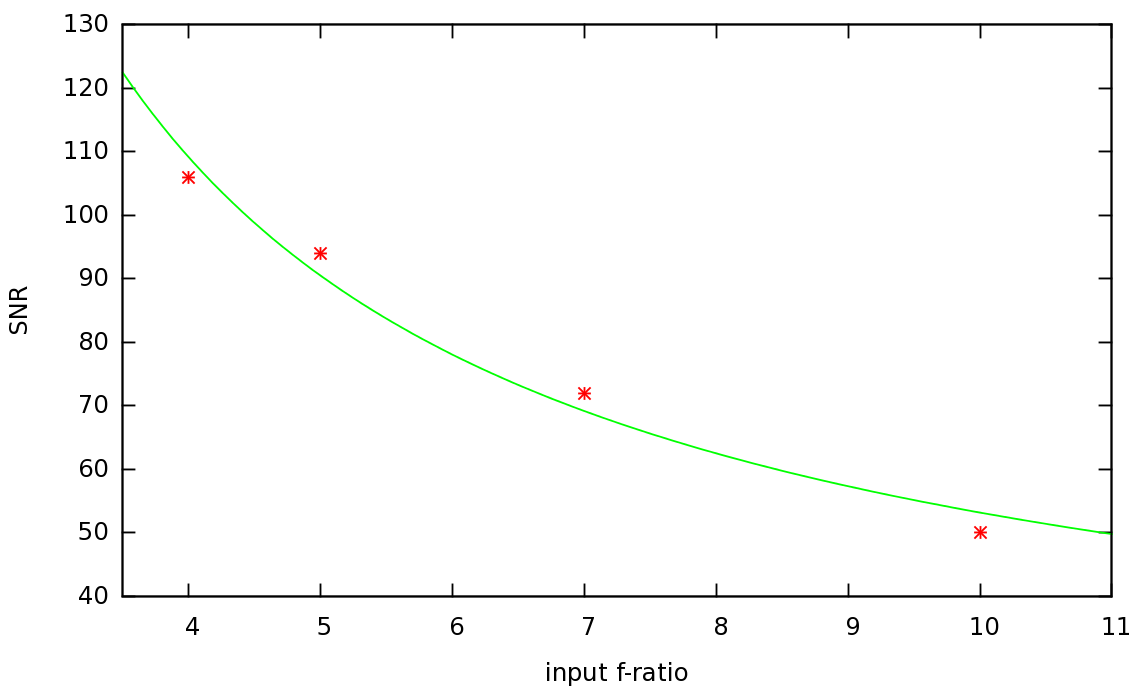}
\end{center}
 \caption{Example of SNR measurements (pluses) and its proportionality to the input f-ratio. The line is a fit of the form  $f(x)=a/F + b$. Fibre (FG050LGA) from Thorlabs.}
  \label{F2}
  \end{figure}

Hence, the number of modes is proportional to $F^{-2}$, where $F$ is the input f-ratio. If we assume that $N_{\rm M}$ is a large integer, which allows to apply Poisson statistics, we can write 
\begin{equation}\label{Eq5}
 \sigma \propto \sqrt{N_{M}} \propto V \propto (NA) \approx (2F)^{-1} \ .
\end{equation}
The signal-to-noise ratio (SNR) is thus given by 
\begin{equation}\label{Eq6}
{\mathrm {SNR}} = N_{\rm M}/\sqrt{N_{\rm M}} \propto F^{-1} \ .
\end{equation}
The SNR is inversely proportional to the f-ratio and we see that modal noise is reduced for smaller wavelengths ($\propto 1/\lambda$), larger fibre cores ($\propto \rho$) and greater input angles ($\propto (NA_{in})$)\footnote{One has to distinguish between the limiting $NA$ given by the fibre materials and $NA_{in}$ of the input cone.}. 
A plot of SNR values {\bf (at around 680 nm, procedure described in Sect. 3.1.)} obtained with a 50 $\mu$m circular fibre from Thorlabs (FG050LGA) is shown in Fig.~\ref{F2} together with a fit of the form $f(x)=a/F + b$, which is what we expect from Eq. (\ref{Eq6}). 

\begin{figure}[th]
\begin{center}
\includegraphics[width=65mm]{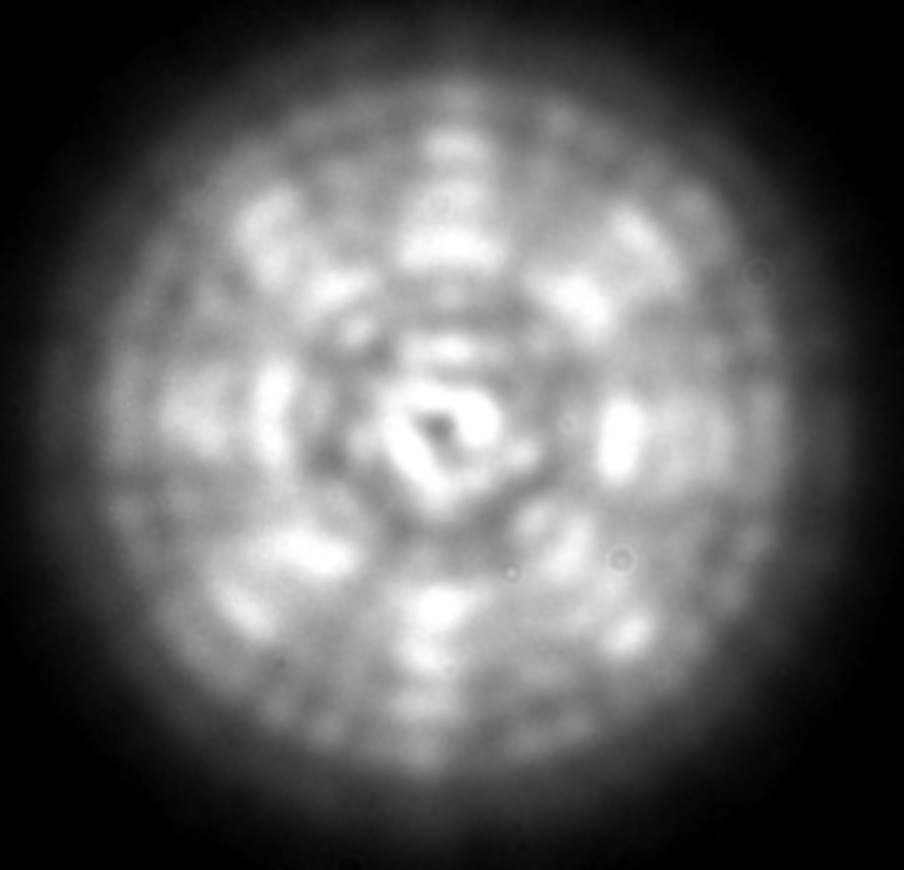}
\end{center}
\caption{Far-field example of a 67 $\mu$m octagonal fibre (WF type from CeramOptec) fed by a microlens {\bf to reimage the pinhole to the fibre entrance}. Visible are eight maxima and minima in azimuthal direction.}
  \label{F3}
\end{figure}

In high-resolution spectroscopy image slicing is an important strategy to achieve the necessary resolution. Since image slicing, acts like a spatial filter, lesser fibre modes take part in image formation\footnote{To be more precise: Since the intensity distribution of modes is rotational symmetric, not a whole mode is vignetted but a part of each mode.} and the effect of modal noise is worse. However, this is, in first place, not an issue related to the fibre used. It is an issue of instrument design, when using a slit or an optical slicer to achieve the necessary resolution. 
Furthermore, the effect is even stronger for bad image quality of the spectrograph because it emphasizes the far-field structure in the focused image.
To illustrate these statements we show a monochromatic far-field image in Fig.~\ref{F3}. This image shows the far-field of a 67 $\mu$m octagonal fibre (WF type from CeramOptec), which was glued to a microlens for direct imaging on the fibre entrance. It shows a clear and well defined symmetric mode structure. Furthermore, eight strong maxima and minima are visible in the azimuthal direction, related to the octagonal structure of the fibre cross-section. Since a spectrograph creates monochromatic images of the input aperture, any effect or technique which expands the spectrum or vignettes modes will further increase the contrast of that structure, i.e., the distance between neighboring interference maxima and minima. This, finally, causes a larger noise amplitude in the measurements.
We further should note here that this effect needs more attention for high-precision measurements of line profiles, since the amplitude changes from observation to observation.

The technique to reduce modal noise is called scrambling and is realized by an optical (e.g. Barnes \& MacQueen \cite{bar:mac}) or a mechanical device. A mechanical scrambler changes the curvature of the fibre, which causes a change of the interference conditions inside the waveguide. This means that the angles $\Theta_{m}$ in Eq. (~\ref{Eq2}) are changing and with that the angular position of the interference maxima in the far-field of the fibre. It is because of moving the fibre by the mechanical scrambler, i.e., the far-field structure is being averaged (homogenized). Since this far-field structure is responsible for the modal noise phenomena, such a homogenization technique is used to reduce modal noise and is called scrambling.
\begin{figure*}[th]
 \begin{center}
 \includegraphics[width=170mm]{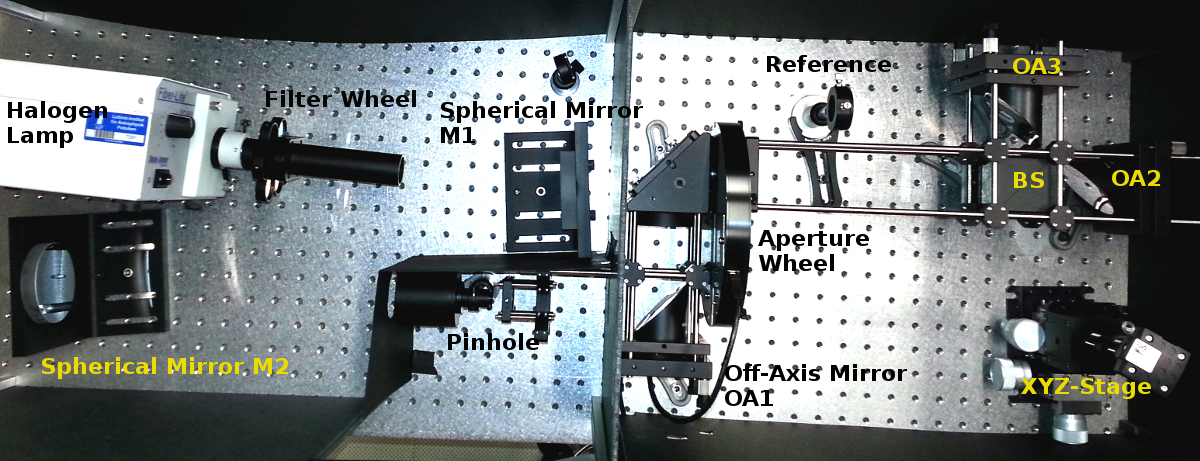}
\end{center}
\caption{Our fibre measurement apparatus for measuring FRD, modal noise, fibre efficiencies and throughput. The pinhole is illuminated by a broadband light source equipped with an filter wheel. An off-axis parabolic mirror (OA1) collimates the beam from the pinhole before the beam enters a beam splitter (BS). The beam diameter is controlled with an aperture wheel. The two beams from the beam splitter are focused by two off-axis parabolic mirrors (OA2 \& OA3.}
  \label{F4}
  \end{figure*}
There are two different phenomena related to fibre scrambling  (see, e.g., Avila~\cite{avila}, Chazelas et al.~\cite{chaz}). (1) The above mentioned modal noise and (2) the stability of the point-spread function (PSF) of the spectrograph. Furthermore, there are differences in dependence of the cross-section geometry. Since this is an intrinsic behavior of the fibre itself, we do not refer to the change of the cross-section geometry as a scrambling technique. In this paper, we do not consider PSF stability but focus on the influence of scrambling on modal noise and focal-ratio-degradation (FRD).

Due to imperfectness of the core-cladding interface the angle which defines the input cone, is not conserved inside the waveguide and is increased at the output. This causes a degradation of the focal ratio at the output compared to the input and is defined as FRD. A curvature of a fibre further decreases the angle of the incident ray at the total reflection between core and cladding (angle $\Theta$ in Fig.~\ref{F1}) and consequently causes an increase of the angle at the fibre exit. This in turn degrades the exit f-ratio with respect to the input f-ratio.

\section{Fibre measurement apparatus}

To obtain repeatable measurements it is particularly important to use a reliable and flexible measurement set-up. Our fibre measurement apparatus (FMA) was designed to address the special needs to characterize optical fibres. The system is adaptable to measure FRD, modal noise, throughput and efficiency of fibres  and fibre-coupling systems. All measurements can be performed as a function of wavelength and input f-ratio. A picture of the whole system is shown in Fig.~\ref{F4}. 

\begin{figure}[th]
 \begin{center}
 \includegraphics[width=85mm]{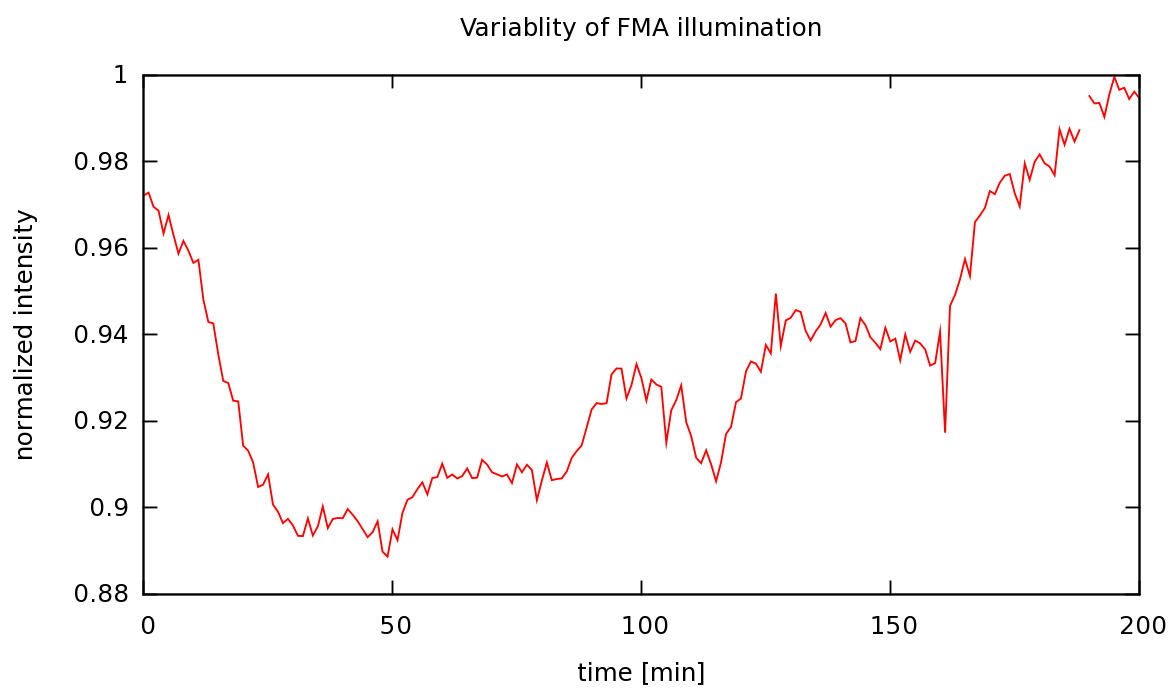}
\end{center}
\caption{Measured fluctuations of the input light source of our fibre measurement apparatus. The long term  ($\approx$ 3\,h) stability is in the range of 5\%, while the short time ($\approx$ min) stability is 0.5\%. To account for lamp-internal thermal effects, the light source is activated one hour before the measurement.}
  \label{F5} 
\end{figure}

\begin{figure}[th]
 \begin{center}
 \includegraphics[width=85mm]{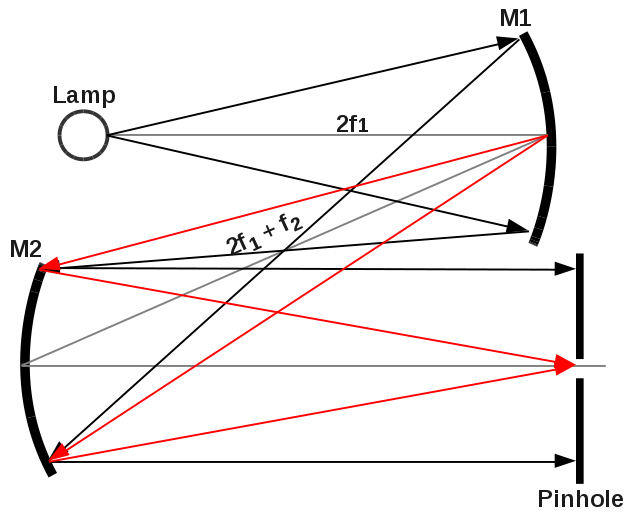}
\end{center}
\caption{Layout of the Koehlerian illumination as used with FMA. The black path shows the light from the lamp, which is parallel at the pinhole position. The pinhole is placed at that very position, where the image of M1 appears (red path).}
  \label{Koehler} 
\end{figure}

A reflective Koehlerian illumination system (shown left in Figs.~\ref{F4} and~\ref{Koehler}) with a Tungsten lamp, adapted from microscopy, is used to illuminate the entrance pinhole of the system. It provides a flat illumination field at the pinhole position. The system creates an image at infinity of the light source. The target (pinhole) is placed where the image of the first mirror (M1) is formed by the second mirror (M2). The broadband Tungsten halogen lamp is equipped with a filter wheel, which contains six different interference filters for wavelength dependent measurements. The first off-axis parabola (OA1, f/3) acts as a collimator. To adjust the f-ratio at the input of the fibre, an aperture wheel is placed within the collimated beam. A beam-splitter (BS, 50:50) creates two beams, one beam to feed the fibre and the other beam for reference.
The reference beam is used to record inherent intensity fluctuations if fibre efficiencies are measured. It is also used to feed a reference fibre in case of measurements of the spatial stability of the fibre output (PSF). This allows to distinguish between fibre and instability effects of the FMA itself. The intensity fluctuations of the light source were measured once and the results are shown in Fig.~\ref{F5}. These fluctuations are recorded with a CCD at the reference beam if necessary (e.g., for transmission and efficiency measurements) and used to correct the measurements for it. A light-source was used with a current control specified by the manufacturer to be stable to 0.5\%, which results in long term intensity fluctuations of 5\%.
To account for thermal effects, the light source was activated one hour before the measurement in order to allow the system to heat-up and stabilize.
A second beam-splitter arrangement was used in front of the fibre to make sure that the image of the entrance pinhole is precisely placed on the fibre core. The fibre itself is mounted to a $xyz$-translation stage for precision positioning. Two off-axis mirrors provide a diffraction limited image on the optical axis without any chromatic aberrations. This is important to eliminate any wavelength dependent errors created by the measurement system itself.

FRD measurements were performed for different wavelengths and for an encircled energy of 95\%. The diameter of the output cone was measured at two different
distances from the CCD. This yields the output f-ratio
\begin{equation}\label{Eq7}
 F_{\rm out}^{95}=s/(D_{1}-D_{2}) = s/d \ ,
\end{equation}
where $D_{1}$ and $D_{2}$ are the diameters of the cone at two distances and $s$ is the longitudinal distance between the measurements. The error is given by the uncertainties of measuring $s$ and $d$ 
\begin{equation}\label{Eq8}
 \sigma_{F}=\sqrt{(\sigma_{s}^{2}/d^{2}+s^{2}\sigma^{2}_{d}/d^{4})} \ .
\end{equation}

\section{Analysis}

All fibres measured are listed in Table~\ref{tab1} for reference. {\bf We note here, that the edges of all the non-circular fibres are rounded and not fully sharp.} As mentioned in the introduction bending a fibre could cause an increase of FRD. Bending the fibre, or more precise, changing the curvature of the fibre with time, is exactly what our scrambler does. Hence, it is important to study FRD behavior of non-scrambled (Sect 3.1) and scrambled (Sect. 3.2) fibres.

\begin{table}
\caption{Fiber identifications for measurements in this paper. Manufacturers are TL Thorlabs, PM Polymicro, CO CeramOptec, and JF J-Fibres. $d$ diameter of the fibre, $L$ length of the fibre.}
\begin{tabular}{lllll}
\hline  \noalign{\smallskip}
No. & Type & Cross    & $d$          & $L$\\
      &         & section & ($\mu$m) & (m)\\
 \noalign{\smallskip}\hline  \noalign{\smallskip}
1 & TL FG050LGA & circular & 50 & 5\\
2 & PM  FBP & circular & 100 & 2\\
3 & PM  FBP & octagonal & 100 & 2\\
4 & PM  FBP & circular & 200 & 2\\
5 & PM  FBP & octagonal & 200 & 2\\
6 & PM  FBP & circular & 300 & 2\\
7 & PM  FBP & octagonal & 300 & 2\\
8 & CO  WF & square & 70 & 2\\
9 & CO  WF & square & 140 & 2\\
10 & CO WF & rectangular & 150x300 & 2\\
11 & JF  NCS & rectangular & 55x150 & 4\\
12 & PM FBP & circular & 200 & 44\\
13 & PM FBP & octagonal & 200 & 44\\
\noalign{\smallskip} \hline
\end{tabular}
\label{tab1}
\end{table}
  
\subsection{Non-scrambled measurements}

We assume that the total noise of our modal noise (SNR) measurements consists of three major sources: (1) background noise $B$ from all possible sources of the measurement system, e.g., like CCD readout noise, (2) photon noise $P$ as the square root of the actual signal $S$ in the spectrum, i.e., $P =\sqrt{S}$ and, (3) modal noise $M$ as the interference structure in the continuum due to the fibre modes. The total noise $N$ is then given by
\begin{equation}\label{Eq9}
 N^{2}=B^{2}+P^{2}+M^{2} \ .
\end{equation}
Because the values for $N$, $P$ and $B$ can be extracted from the CCD readout, the modal noise is then calculated via
\begin{equation}\label{Eq10}
 M=\sqrt{N^{2}-B^{2}-S} \ .
\end{equation}

The background is measured on a part of the CCD frame where no signal from the spectrum is present and is assumed to be constant over the surface of the CCD array. To reduce the background noise, and thus the error introduced to the measurements, ten CCD frames are combined and the median of the frames used. A Czerny-Turner laboratory spectrograph with a Moravian G2-1600 CCD was used for the measurements. This spectrograph suffers from a great amount of astigmatism, which causes the spectrum to expand several times (perpendicular to the direction of dispersion) as geometrically expected. This provides good visibility of the mode structure in the continuum spectrum of the used broadband (halogen) lamp. All measurements in this subsection are done at the same wavelength range between 674\,nm and 698\,nm.
The resolution of the spectrograph depends on the fibre core diameter. For a 50 $\mu$m core the theoretical resolving power, $\lambda/\Delta\lambda$, of the spectrograph is $\approx$ 20,000 and the dispersion is around 0.0157\,nm/pix at 9\,$\mu$m pixel size.

\begin{figure}[h!]
 \begin{center}
 \includegraphics[width=85mm]{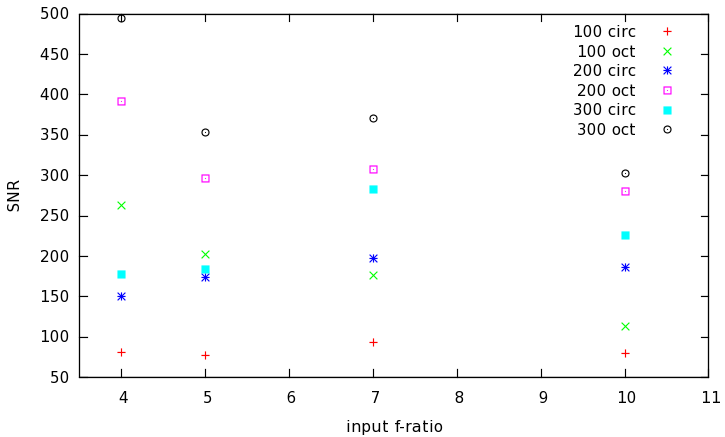}
\end{center}
\caption{Measured SNR for circular and octagonal fibres. The core diameters and cross-section geometries  are listed in the insert at the top right.}
  \label{F6}
  \end{figure}
  
Measurements for fibres with circular and octagonal cross-sections are shown in Fig.~\ref{F6}. The measured SNR for, e.g., an f/10 input beam can be grouped like the following; 100 $\mu$m circ $\textless$ 100 $\mu$m oct $\textless$ 200 $\mu$m circ $\textless$ 300 $\mu$m circ $\textless$ 200 $\mu$m oct $\textless$ 300 $\mu$m oct. As expected, the larger the fibre core the better the SNR. A larger fibre core causes more modes to be excited inside the waveguide and since these modes are spatially separated but still imaged onto the same area on the CCD, the distance between the modes is smaller, i.e., the structure of the continuum due to that interference shows a smaller contrast,
and thus lower noise amplitude. For any of the three core diameters the octagonal-shaped fibres enable a higher SNR than their circular counterpart. The measured SNR values listed in Table \ref{tab2} are {\bf for f/7} input f-ratio for a given fibre and show a clear increase of SNR with increasing fibre-core diameter. 

\begin{table}
\caption{Measured SNR values for circular and octagonal fibres for an input f-ratio of 7 in dependence of the fibre-core diameter.}
\begin{tabular}{lllll}
\hline  \noalign{\smallskip}
core   & averaged SNR & averaged SNR \\
size & for circular & for octagonal \\
 \noalign{\smallskip}\hline  \noalign{\smallskip}
100 & 93 & 177\\
200 & 197 & 308\\
300 & 283 & 371\\
\noalign{\smallskip} \hline
\end{tabular}
\label{tab2}
\end{table}

Dividing the SNR values of the octagonal fibres by those of the circular fibres and averaging over the core sizes yields an average factor for the improvement of the SNR in dependence of the input f-ratio. This is listed in Table~\ref{tab3} for the four measured input f-ratios (second column). Furthermore, the factors in dependence of the fibre-core size computed in an equivalent way are listed in the same table (column 4).

The result is more pronounced at an f/4 input where we recorded the following grouping; 100 $\mu$m circ $\textless$ 200 $\mu$m circ $\textless$ 300 $\mu$m circ $\textless$ 100 $\mu$m oct $\textless$ 200 $\mu$m oct $\textless$ 300 $\mu$m oct. In this case, even the 100 $\mu$m octagonal fibre suffers less from modal noise than the 300 $\mu$m circular fibre. 

\begin{table}
\caption{Improvement of measured SNR values for octogonal over circular fibers. Left: averaged over the fibre-core diameters for the four measured input f-ratios. Right: fibre-core diameters averaged over the input f-ratios.}
\begin{tabular}{cc|cc}
\hline  \noalign{\smallskip}
input   & averaged & core-size & averaged\\
f-ratio & factor & $\mu$m & factor\\
 \noalign{\smallskip}\hline  \noalign{\smallskip}
4 & 2.87 & 100 & 2.3\\
5 & 2.09 & 200 & 1.83\\
7 & 1.59 & 300 & 1.84\\
10 & 1.43\\
\noalign{\smallskip} \hline
\end{tabular}
\label{tab3}
\end{table}

A closer look at Fig. \ref{F6} reveals some peculiarities. The values for f/7 input are for most of the measured fibres higher than the ones for f/5 input, which indicates a local maximum. Furthermore, the values for all the circular fibres have their global maximum at f/7 and not at fast input ratios as expected. A more detailed study of this effect would be interesting, but is out of the scope of this work.

\begin{figure}[th]
 \begin{center}
 \includegraphics[width=85mm]{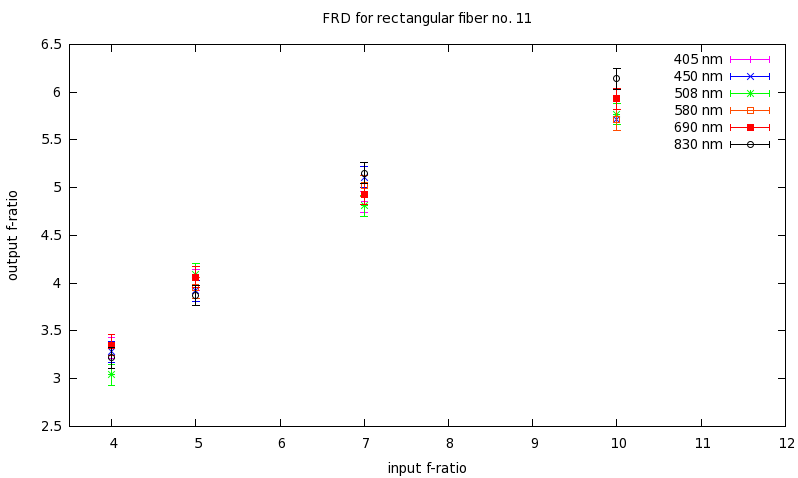}
\end{center}
  \caption{Measured FRD values for the rectangular fibre no. 11 from J-Fiber (see Table~\ref{tab1}).}
  \label{F7}
  \end{figure}

\begin{figure*}[th]
 \begin{center}
 \includegraphics[width=170mm]{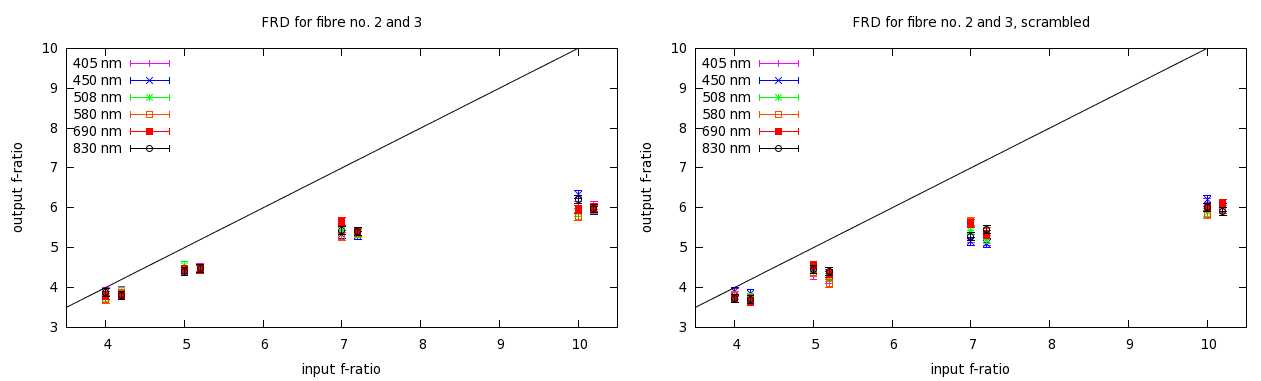}
\end{center}
\caption{Measured FRD values (right: non-scrambled, left:scrambled) for 100 $\mu$m circular (shifted by + 0.2 in x for visibility) and octagonal fibre no.~2 ans 3 from Polymicro (see Table~\ref{tab1}).}
  \label{F9}
  \end{figure*}

\begin{figure*}[th]
 \begin{center}
 \includegraphics[width=170mm]{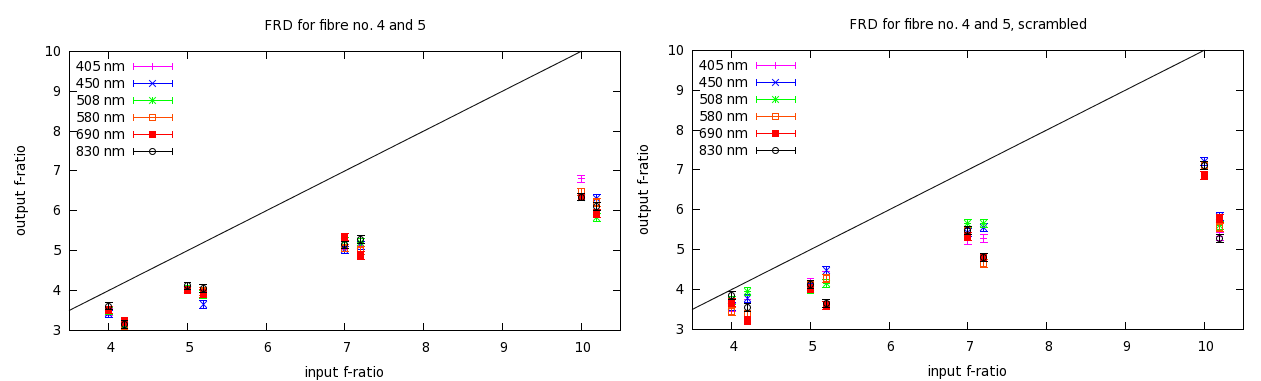}
\end{center}
  \caption{Measured FRD values (right: non-scrambled, left:scrambled) for 200 $\mu$m circular (shifted by + 0.2 in x for visibility) and octagonal fibre no. 4 and 5 from Polymicro (see Table~\ref{tab1}).}
  \label{F10}
  \end{figure*}

\begin{figure*}[th]
 \begin{center}
 \includegraphics[width=170mm]{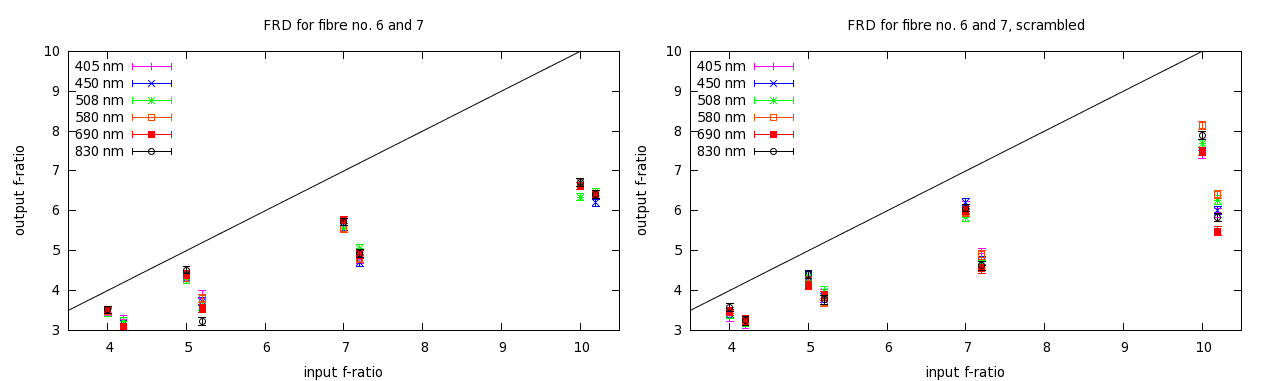}
\end{center}
  \caption{Measured FRD values (right: non-scrambled, left:scrambled) for 300 $\mu$m circular (shifted by + 0.2 in x for visibility) and octagonal fibre no. 6 and 7 from Polymicro (see Table~\ref{tab1}).}
  \label{F11}
  \end{figure*}
  
 \begin{figure}[th]
 \begin{center}
 \includegraphics[width=85mm]{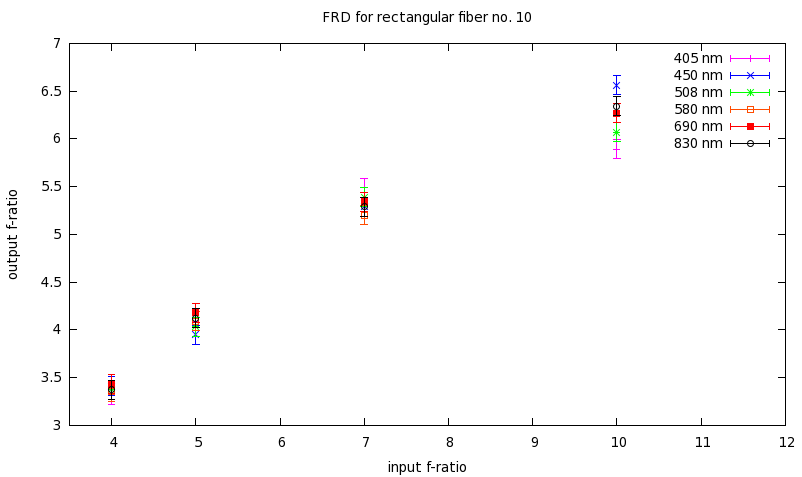}
\end{center}
  \caption{Measured FRD values for rectangular WF fibre no. 10 from CeramOptec (see Table~\ref{tab1}).}
  \label{F8}
  \end{figure}

\begin{figure}[th]
 \begin{center}
 \includegraphics[width=85mm]{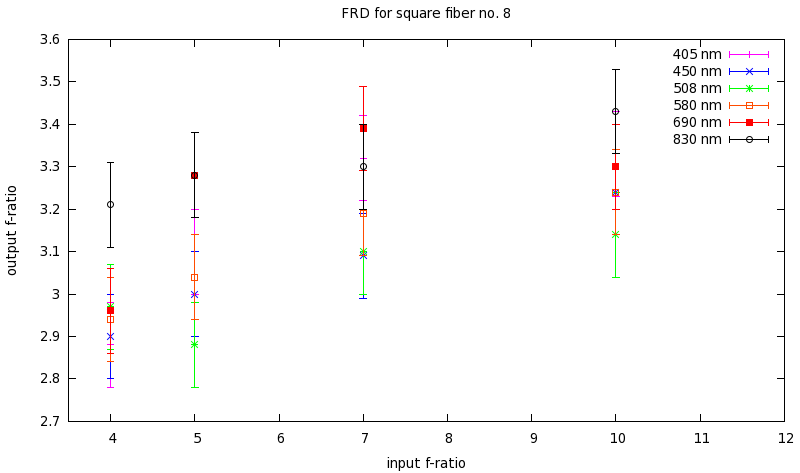}
\end{center}
  \caption{Measured FRD values for the square WF fibre no. 8 from CeramOptec (see Table~\ref{tab1}). This fibre shows worse FRD behavior, even an input of f/10 yields a fast output f-ratio around 3.2.}
  \label{F12}
  \end{figure}

\begin{figure}[th]
 \begin{center}
 \includegraphics[width=85mm]{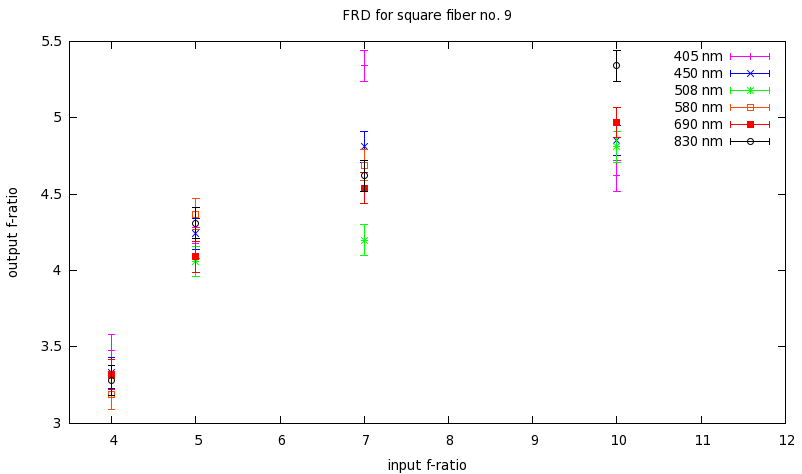}
\end{center}
  \caption{Measured FRD values for the square WF fibre no. 9 from CeramOptec (see Table~\ref{tab1}). This fibre shows for fast input f-ratio of 4 similar values as the rectangular fiber no. 11 in Fig.~\ref{F7}, but for a slow input of f/10 it yields a faster output.}
  \label{F13}
  \end{figure}
  
The FRD values for circular and octagonal fibres are shown in Figs.~\ref{F9} -\ref{F11}. For fibre no.~11 (rectangular NCS fibre) the FRD values are shown in Fig.~\ref{F7}. These are comparable to the values obtained for the circular 200 $\mu$m fibre no. 4 in Fig.~\ref{F10}.
Octagonal fibres show a slightly better FRD behavior than the corresponding circular fibres. 
This can be seen in Fig. \ref{F10} (fibres no. 4 and 5) where the circular fibre provides an output f-ratio of 3.2 and the octagonal fiber an output f-ratio of 3.5, both for an input f-ratio of 4. 
The values for all measured circular and octagonal fibres are summarized in Table~\ref{tab4}. Fig.~\ref{F12} and Fig.~\ref{F13} show the FRD values for the square fibres no.~8 and 9, respectively. These suffer from worse FRD behavior and strong scatter. This is pronounced in the measurements shown in Fig. \ref{F12} for fibre no. 8, were the output f-ratio is around 3.3 even for an input f-ratio of 10. 
Furthermore, it seems that green light (508 nm) yields the fastest output, while red (690 nm) and IR (830 nm) light yields the slowest output. However, one fibre sample is not enough to make definite conclusions for wavelength dependencies. For fibre no. 9 in Fig. \ref{F13}, we measured an output f-ratio around 5 for an input f-ratio of 10. These results were still the same 
after three repetitions with the same sample fibre. It can not be said that this behavior is limited to squared fibres because we only measured one sample for each of the two types of squared fibres.

\begin{figure}[h!]
 \begin{center}
 \includegraphics[width=85mm]{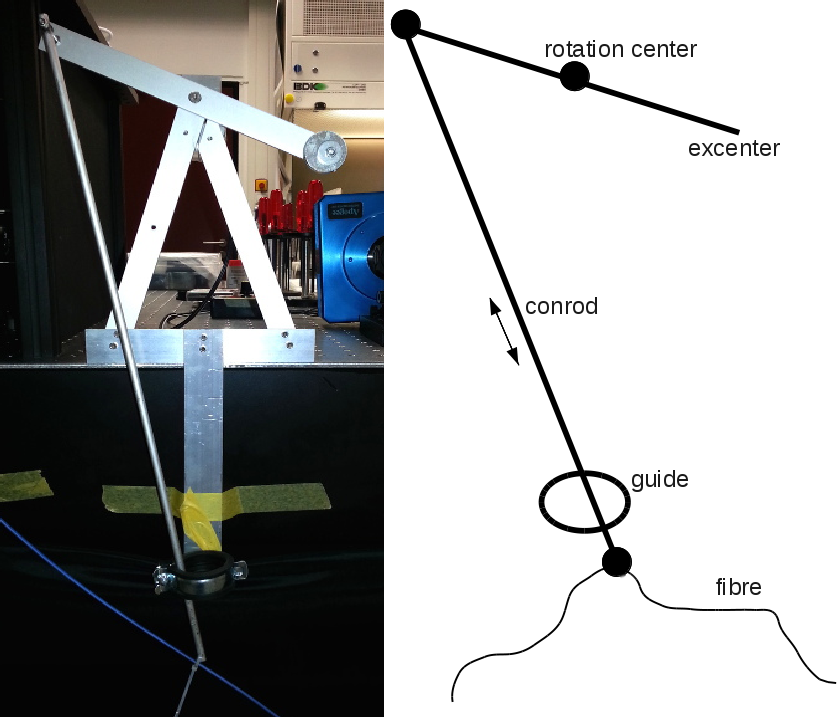}
\end{center}
  \caption{Excenter as a simple mechanical scrambler. The fibre is connected to the pleuel which itself is connected to the rotating excenter. The rotation frequency is adjustable between 50 and 300 per minute.}
  \label{F14}
  \end{figure}

\begin{figure}[h!]
 \begin{center}
 \includegraphics[width=85mm]{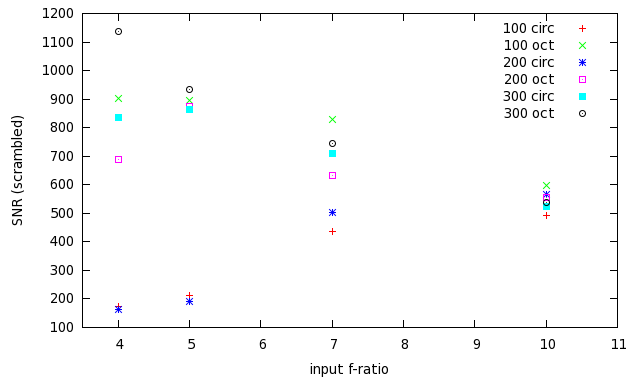}
\end{center}
  \caption{Measured SNR for circular and octagonal fibres with mechanical scrambling. Note that the smaller circular fibers deviate from the expected trend.}
  \label{F15}
  \end{figure}

\subsection{Scrambled measurements}

In order to increase the SNR, we try to reduce the amount of modal noise by a simple scrambling method. The scrambler is purely mechanical and is shown in Fig.~\ref{F14}. The fibre to be tested is connected to the excenter (with amplitude of 20 cm) of the scrambler. The fibre is fixed on both sides of the scrambler to prevent
vibrations introduced by the shaked fibre itself. The rotation frequency of the excenter is adjustable between 50 and 300 rotations per minute. For an exposure time of two seconds a frequency of $\nu_{s}$ = 120 per minute was used, which gives four rotations within the exposure time. The critical exposure time is  $t_c$ = 0.5\,sec in order to cover at least one half of a rotation (full amplitude) of the excenter.

Because a mechanical scrambler physically bends the fibre to change the interference conditions within the fibre core, it is important to make sure that a scrambler does not increase any other degrading factors e.g., FRD or transmission losses. Therefore, FRD is measured also with scrambling applied. By comparing these measurements 
for non-scrambled with the scrambled measurements shown in Fig. \ref{F9} to \ref{F11} it can be seen that this high-amplitude low-frequency scrambler does not change the FRD behavior of the fibres. Table~\ref{tab4} lists the measured output f-ratios for an input f-ratio of 4 averaged over the different wavelengths.

\begin{table}
\caption{Measured output f-ratios for corresponding circular and octagonal fibres for input f-ratio 4. There is no significant increase of the output for the scrambled fibres. The values are averaged over the wavelengths.}
\begin{tabular}{lllll}
\hline  \noalign{\smallskip}
fibre   & output f-ratio & output f-ratio\\
type    & non-scrambled & scrambled \\
\noalign{\smallskip}\hline  \noalign{\smallskip}
100 $\mu$m oct & 3.8 $\pm$ 0.1 & 3.8 $\pm$ 0.1 \\
100 $\mu$m circ & 3.8 $\pm$ 0.1 & 3.7 $\pm$ 0.1\\
200 $\mu$m oct & 3.5 $\pm$ 0.2 & 3.6 $\pm$ 0.1\\
200 $\mu$m circ & 3.1 $\pm$ 0.1 & 3.5 $\pm$ 0.2\\
300 $\mu$m oct & 3.5 $\pm$ 0.1 & 3.5 $\pm$ 0.1\\
300 $\mu$m circ & 3.1 $\pm$ 0.1 & 3.2 $\pm$ 0.1\\
\noalign{\smallskip} \hline
\end{tabular}
\label{tab4}
\end{table}

\section{Tests with long science fibres}

By the end of February 2013 we replaced the 50 $\mu$m circular fibre of the STELLA-\'echelle-spectrograph (SES) (Weber et al.~\cite{SES}) with a 50 $\mu$m octagonal fibre from CeramOptec. To determine the gain in SNR, we measured the line-poor star \object{Vega} with both fibres and determined the noise defined as the standard deviation of the signal within a certain wavelength range. {\bf We applied the same procedure as described in Sect. 3.1.}
The measured values are listed in Table \ref{tabSES}, where we can clearly see a reduction of noise for the octagonal fibre for all wavelengths.
The small SNR values for this bright star are due to the fact that we lowered the exposure time to get rid of any scrambling caused by the movement of the fibre while the telescope is tracking.
We further have corrected these values for a difference in the exposure level (difference of photon noise) by a factor of 1.4, i.e. the spectrum with the circular fibre has had a higher signal. 

\begin{table}
\caption{Measured SNR with the Stella-\'Echelle-Spectrograph for circular and octagonal 50 $\mu$m fibre for different wavelengths in the continuum of Vega spectra.}
\begin{tabular}{lllll}
\hline  \noalign{\smallskip}
wavelength $\lambda$  & circular fibre & octagonal fibre & ratio SNR\\
nm & SNR & SNR & oct/circ \\
\noalign{\smallskip}\hline  \noalign{\smallskip}
433.1 & 20 & 27 & 1.35\\
649.4 & 31 & 49 & 1.58\\
666.8 & 48 & 63 & 1.31\\
678.1 & 73 & 117 & 1.60\\
\noalign{\smallskip} \hline
\end{tabular}
\label{tabSES}
\end{table}

Furthermore, we did some similar SNR measurements as in Sect. 3 with the long science fibres for PEPSI. These fibres are also from Polymicro's FBP family and have a length of 44 m (Strassmeier et al.~\cite{pepsi}). If we compare these with the ones shown in Fig.~\ref{F6} for the unscrambled short fibres, we see some interesting differences. Firstly, the local maxima at around f/7 is no more clearly visible.
Secondly, the increase of SNR for lower f-numbers is still visible and supports our previous measurements. If we take the imperfectness of the core-cladding interface as a source of intrinsic scrambling, the SNR should increase with the number of reflections the wave undergoes within the waveguide.
We can not support this from our measurements. There is an indication that smaller fibres (which suffer from more reflection for a given length of the waveguide) have the greatest increase of SNR, but more fibres need to be measured. However, the 100 $\mu$m
circular fibre still shows the smallest SNR, but increased significant. A similar situation is present for the 100 $\mu$m octagonal fibre, which shows a large increase of the SNR.

\begin{figure}[h!]
 \begin{center}
 \includegraphics[width=85mm]{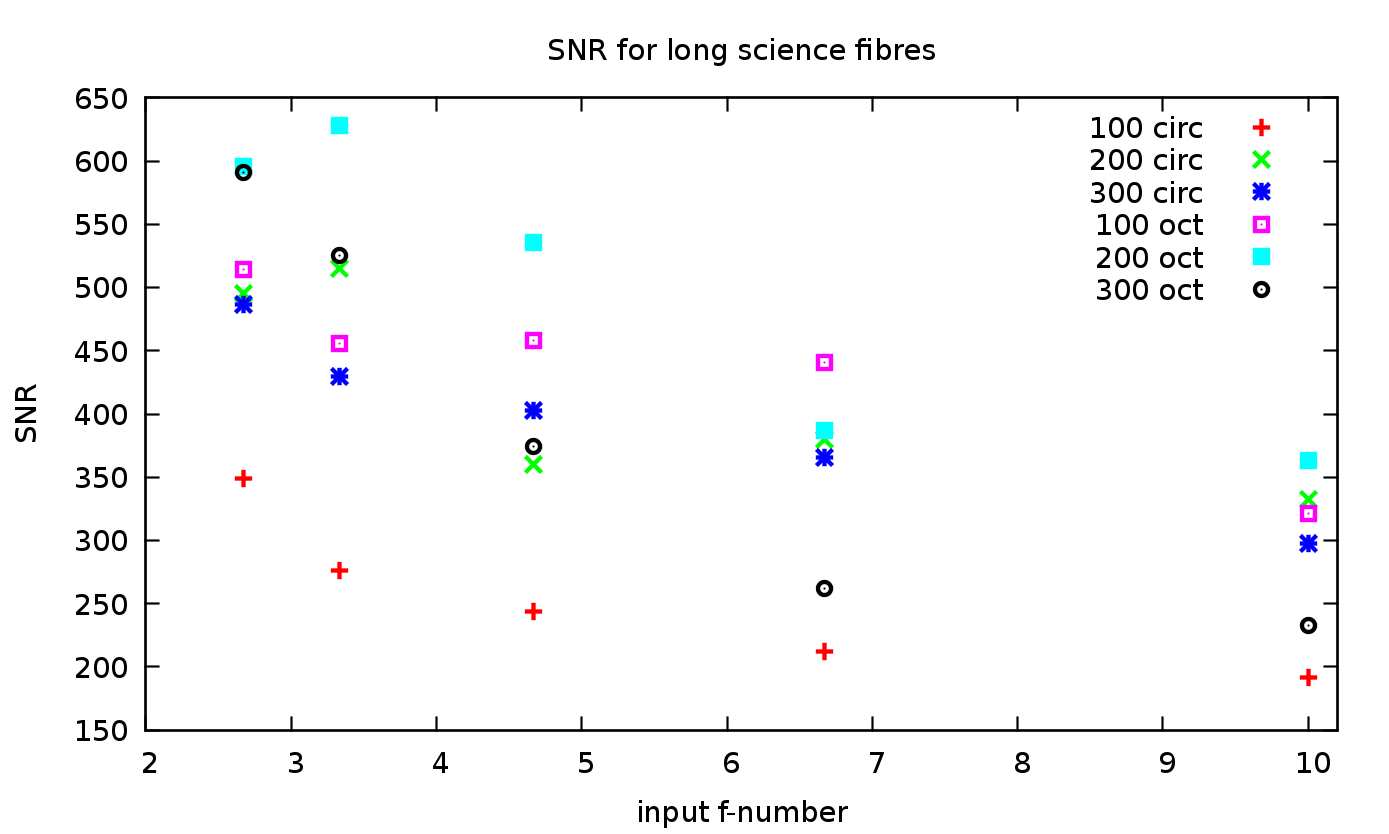}
\end{center}
  \caption{Measured SNR for the long (44 m) science fibres of PEPSI (see text).}
  \label{F16}
  \end{figure}
  
\begin{figure}[h!]
 \begin{center}
 \includegraphics[width=85mm]{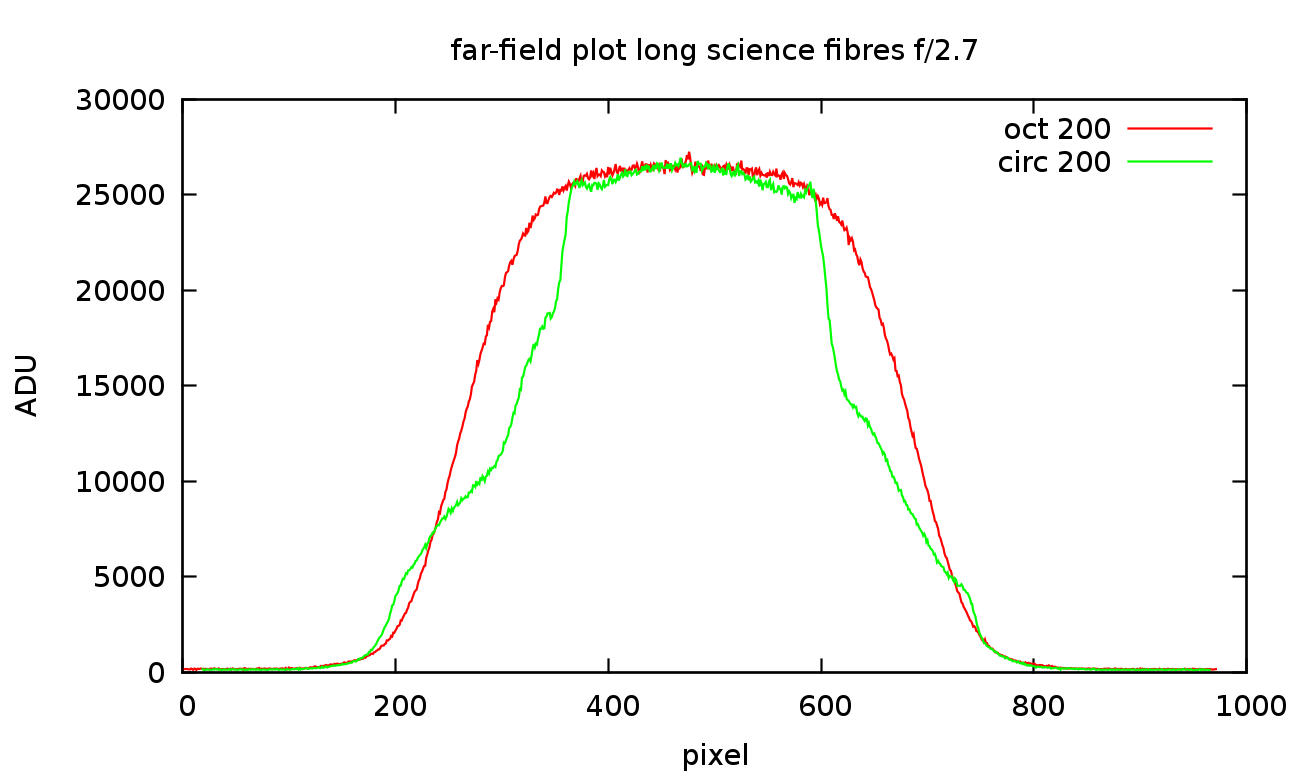}
\end{center}
  \caption{Far-Field for f/2.7 input and a FWHM of the image {\bf of the pinhole} of 50 $\mu$m for 200 $\mu$m fibres.}
  \label{F17}
  \end{figure}

\begin{figure}[h!]
 \begin{center}
 \includegraphics[width=85mm]{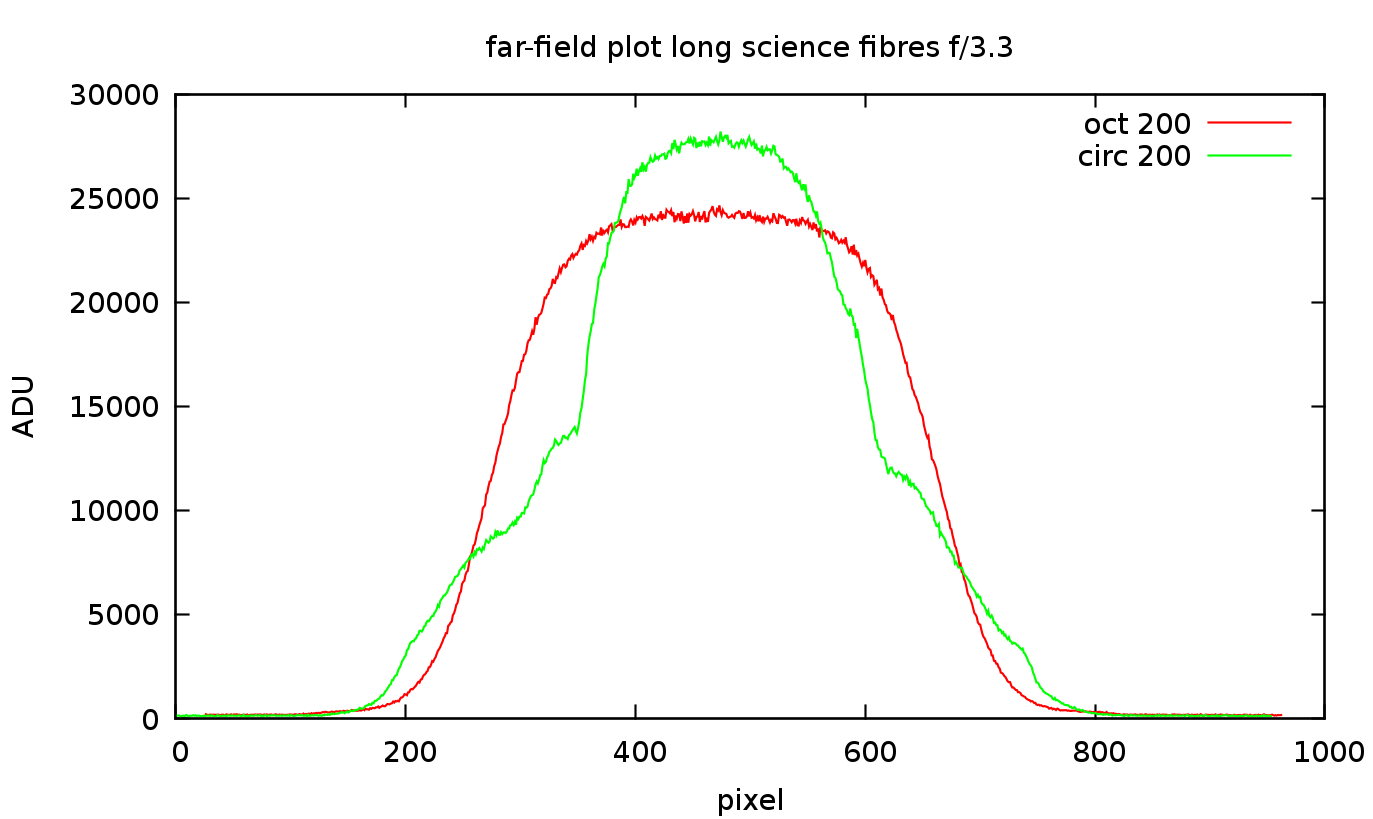}
\end{center}
  \caption{Far-Field for f/3.3 input and a FWHM of the image {\bf of the pinhole} of 50 $\mu$m for 200 $\mu$m fibres.}
  \label{F18}
  \end{figure}

\begin{figure}[h!]
 \begin{center}
 \includegraphics[width=85mm]{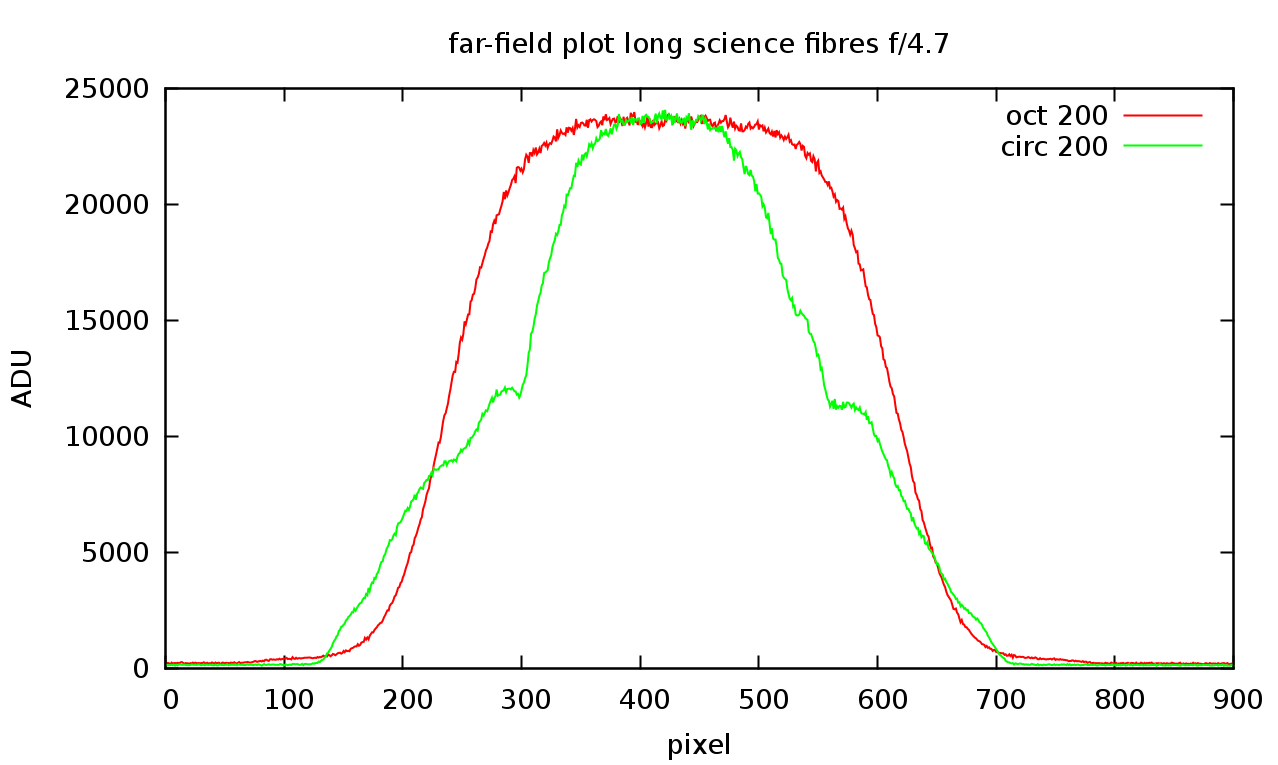}
\end{center}
  \caption{Far-Field for f/4.7 input and a FWHM of the image {\bf of the pinhole} of 50 $\mu$m for 200 $\mu$m fibres.}
  \label{F19}
  \end{figure}

\begin{figure}[h!]
 \begin{center}
 \includegraphics[width=85mm]{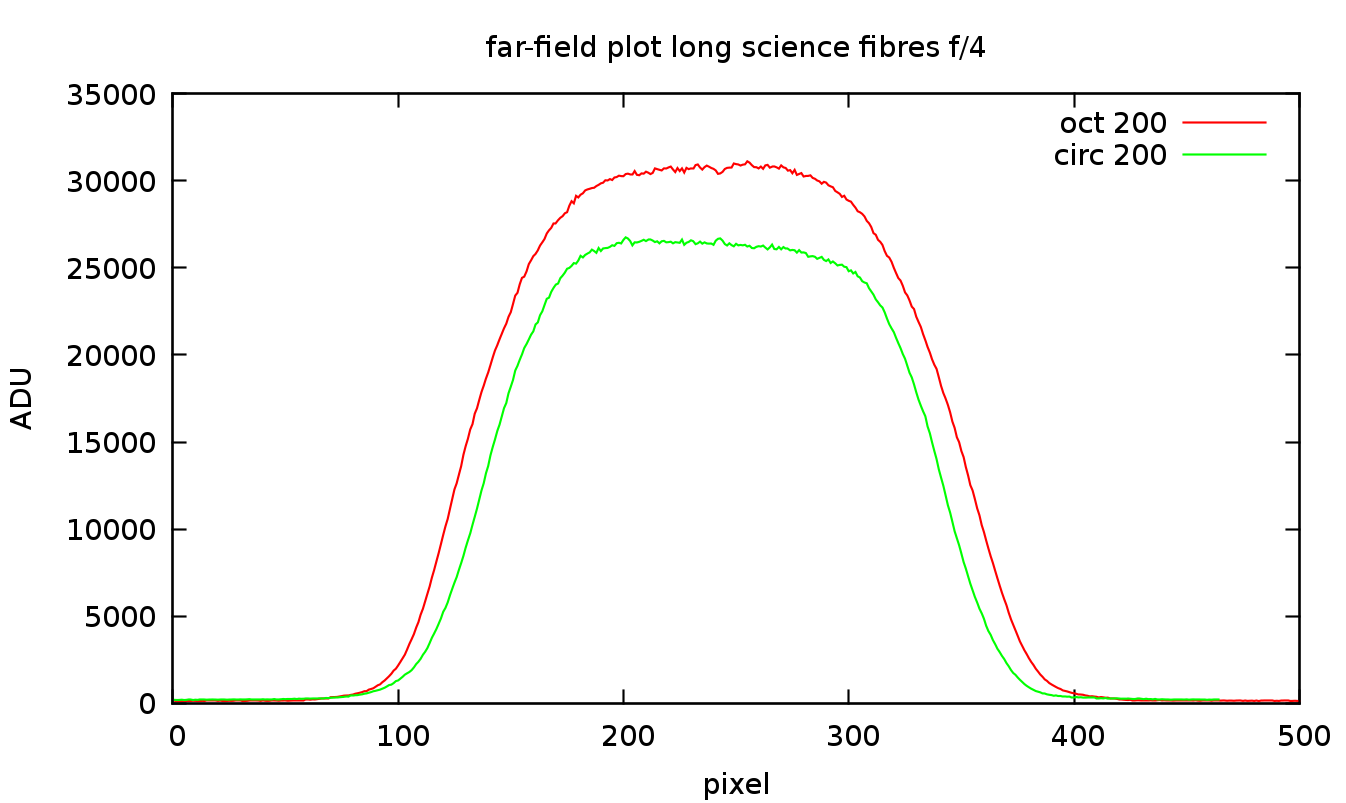}
\end{center}
  \caption{Far-Field for f/4 input and a FWHM of the image {\bf of the pinhole} of 100 $\mu$m for 200 $\mu$m fibres.}
  \label{F20}
  \end{figure}

Additionally to above SNR measurements, we observed the far-field at different input f-ratios and image sizes {\bf of the pinhole at the fibre entrance}. We directly imaged the pinhole of the FMA onto the fibre entrance. We show four examples of these far-fields; Fig.~\ref{F17} (f/2.7 input and 50 $\mu$m FWHM image size), Fig.~\ref{F18} (f/3.3 input and 50 $\mu$m FWHM image size), Fig.~\ref{F19} (f/4.7 input and 50 $\mu$m FWHM image size) and Fig.~\ref{F20} (f/4 input and 100 $\mu$m FWHM image size).
The octagonal fibres show also a better scrambling behavior in the far-field for strong underfilling of the fibre core. Note that for the 50 $\mu$m FWHM only 1/16 of the fibre entrance is illuminated. These cases can become important for pupil imaging, i.e., for seeing dependend input f-ratios at the telescope.

\section{Summary and conclusions}

\begin{figure}[h!]
 \begin{center}
 \includegraphics[width=85mm]{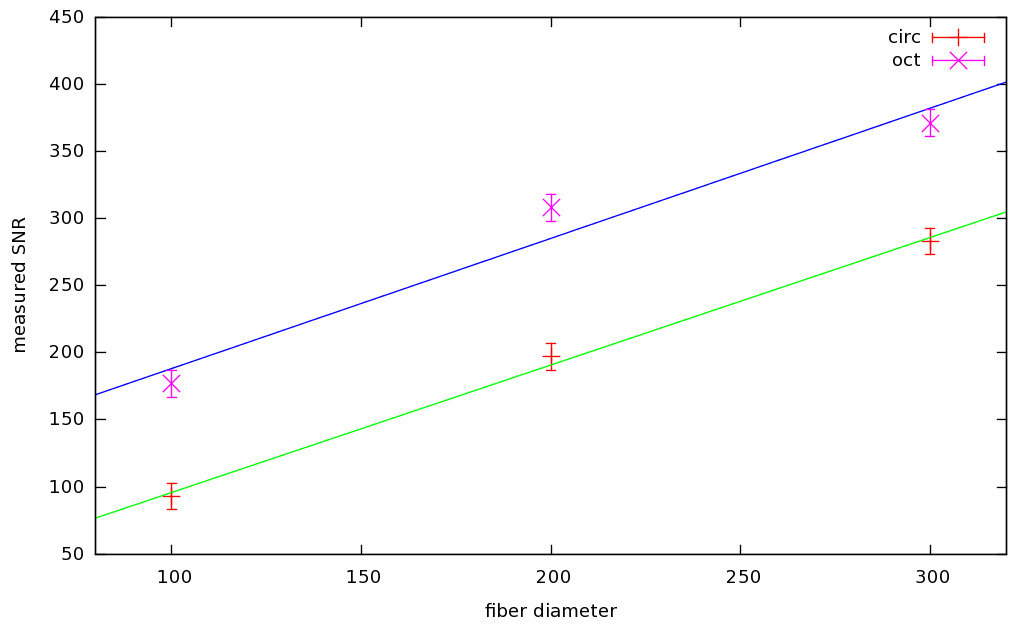}
\end{center}
  \caption{Plotted are the SNR values listed in Tab. ~\ref{tab2} (non-scrambled). The increase of SNR with increasing fibre diameter is clearly visible. Furthermore, we applied a linear fit ($f(x) = ax+b$) to these measurements and list the coefficients in Tab.~\ref{tab5}.}
  \label{F21}
  \end{figure}

We have shown that the FRD and modal noise behavior of octagonal fibres is superior to circular fibres. The values are summarized in Table~\ref{tab4} for FRD and in Table~\ref{tab2} and~\ref{tab3} for modal noise. As indicated in the introduction from a theoretical point of view, modal noise decreases with increasing fibre diameter. This is supported by our measurements as shown in Table~\ref{tab2}. 
If we apply a linear fit ($f(x) = ax+b$) to these measurements, we get the results shown in Fig.~\ref{F21} listed in Table~\ref{tab5}. However, it is necessary to measure fibres with other diameters in order to increase the statistics and prove the linear dependency. 
Furthermore, it should be noted that the fibre parameter $V$ is derived from an electromagnetic description only for rotational symmetric fibres with circular geometry (ref. e.g. Bures \cite{bures}).

\begin{table}
\caption{The parameters for a linear fit ($f(x) = ax+b$) to the SNR measurements listed in Tab.~\ref{tab2} (non-scrambled) of circular and octagonal cross-section fibres.}
\begin{tabular}{lllll}
\hline  \noalign{\smallskip}
fibre type  & a & b\\
\noalign{\smallskip}\hline  \noalign{\smallskip}
circ & 0.95 $\pm$ 0.05 & 1.0 $\pm$ 11.3 \\
oct  & 0.97 $\pm$ 0.20 & 91.33 $\pm$ 42.41\\
\noalign{\smallskip} \hline
\end{tabular}
\label{tab5}
\end{table}

Comparing the non-scrambled measurements in Fig.~\ref{F6} with the scrambled measurements in Fig.~\ref{F15} shows a strong increase of SNR for the scrambled applications. One example is the measurement for fibre no. 7, which yields an improvement of $\approx$ 2 for the applied scrambling. However, the reduction of modal noise is even higher for the smaller core fibres, e.g., fibre no. 3 (100 $\mu$m octagonal) shows an improvement of $\approx$ 3 for f/4 and $\approx$ 5 for f/10 input, respectively. 

Since no geometrical parameter of the fibre changes (bending, thermal expansion / contraction etc.), and also not the illumination, the structure in the output far-field stays the same for all times. We applied a periodic scrambler, which changes the bending of the fibre in a reproducible way, i.e., the output structure is the same for a given phase position of the excenter. As already mentioned, the minimum exposure time, $t_c$, is limited to the maximum frequency of the excenter. 
However, the maximum frequency for our scrambler is around 200 rotations per minute in order to avoid damage of the fibre itself. Therefore, exposure times shorter than $t_c$ lead to single-shot effects, i.e., the far-field is not well homogenized and modal noise is still at a high level. However, this effect is of lower priority because exposure times for high-resolution spectrographs are commonly much longer than our critical time $t_c$. Furthermore, this problem is solvable by averaging many frames obtained at different phase positions. 
Another posibility is to use a non-periodic scrambler. Since such a scrambler is rotating with different speed, it is hard to calculate a critical exposure time and to get reproducible results, which is necessary for characterizing fibres inside the laboratory. 

The presented FRD measurements show that our scrambling method does not influence the FRD performance, neither for circular nor for octagonal or any other (square or rectangular) cross-section fibres. Figure \ref{F9} shows the non-scrambled measurements for FRD of fibre no. 3, which are equal (within the errors) with those for the scrambled measurements. Furthermore, measured FRD values averaged over wavelengths for an input f-ratio of 4 are listed in Table~\ref{tab4} for the circular and octagonal fibres.
It should ne noted that, as for most measurements with fibre optics, the statistics of the measurements should be improved by measuring more fibre samples 
to distinguish between measurement errors and real dependencies, e.g., the wavelength dependence of FRD.

The test at the STELLA-\'echelle-spectrograph has shown an increase of at least a factor of 1.3 (Tab.~\ref{tabSES}) in SNR for the octagonal fibre. As expected, longer fibres also show a better scrambling of the intensity structure at the output, especially for small-core fibres. The trend of increasing SNR with decreasing input f-ratio is also conserved. A further advantage of octagonal fibres is the good far-field illumination property for small image sizes, which becomes important when the fibre is fed via pupil imaging.

From these measurements, we can finally conclude that octagonal fibres are superior in both FRD and modal noise compared to circular fibres. The differences 
are small for the FRD behavior but significant in modal noise. Our measurements show a clear advantage in using octagonal fibres if high precision and high SNR measurements are necessary as it is the case for high-resolution spectroscopy, e.g., in the science cases for PEPSI.

\begin{acknowledgements}

Based (partly) on data obtained with the STELLA robotic telescopes in Tenerife, an AIP facility jointly operated by AIP and IAC.
We thank the referee, Gerardo Avila, for his constructive suggestions which helped to improve the paper.

\end{acknowledgements}

\end{document}